\def\cm{cm$^{-1}$}
\def\efa{EuFe$_2$As$_{2}$}
\def\ts{$T_{\rm SDW}$}
\documentclass[prb,twocolumn,showpacs,amsmath,amssymb]{revtex4}
\usepackage[dvips]{graphicx}
\usepackage{graphicx}

\begin{document}

\title{Effects of the magnetic orderings on the
dynamical conductivity:\\
 optical investigations of \efa\ single crystals}

\author{D. Wu}
\email{dan.wu@pi1.physik.uni-stuttgart.de}
\author{N. Bari\v{s}i\'{c}}
\author{N. Drichko}
\author{S. Kaiser}
\author{A. Faridian}
\author{M. Dressel}
\affiliation{1.~Physikalisches Institut, Universit\"at Stuttgart, Pfaffenwaldring 57, 70550 Stuttgart, Germany}
\author{S. Jiang}
\author{Z. Ren}
\author{L. J. Li}
\author{G. H. Cao}
\author{Z. A. Xu}
\affiliation{Department of Physics, Zhejiang University, Hangzhou 310027, People's Republic of China}
\author{H. S. Jeevan}
\author{P. Gegenwart}
\affiliation{I. Physikalisches Institut,
Georg-August-Universit\"at G\"ottingen, 37077 G\"ottingen,
Germany}

\date{\today}

\begin{abstract}
The magnetic, transport and optical properties of \efa\ single
crystals have been investigated parallel and perpendicular to the
$ab$-plane. The anisotropy $\rho_c/\rho_{ab}\approx 8$ depends only
slightly on temperature. In both orientations, the spin-density wave
transition at $T_{\rm SDW}=189$~K shows up as a considerable
increase in the dc resistivity. Susceptibility measurements
evidence the magnetic order of the Eu$^{2+}$ moments at $T_N=19$~K
with little influence on the electronic transport taking
place in the FeAs layers. Polarization-dependent infrared
spectroscopy reveals strongly anisotropic optical properties and
yields a carrier density of only $4.2\times 10^{21}~{\rm cm}^{-3}$
and a bandmass of $m_b=2m_0$. A sizeable Drude contribution
is present at all temperatures and narrows upon cooling.
Below \ts, the
spin-density-wave gap develops in the in-plane optical
conductivity; no appreciable change is detected for the
perpendicular polarization. Modifications in the phonon features
are associated with changes of the electronic properties at \ts.
The extended Drude analysis yields a linear behavior of the frequency-dependent scattering rate below \ts, indicating an interaction between the charge carriers and spin fluctuations in the spin-density-wave state.
\end{abstract}

\maketitle

\section{Introduction}
There is a small number of fundamental issues in solid state science that remain of central importance
even when a new class of materials emerge.
Besides the effect of reduced dimensions and the Mott transition,
the interplay of magnetism and superconductivity is certainly among those fascinating topics. It has been intensively explored in heavy-fermion compounds,\cite{Ott87} 
organic conductors \cite{Ishiguro98} and cuprates,\cite{Ichikawa00} all of them exhibit certain types of magnetic order in the close vicinity to a superconducting phase of unconventional nature.

The observation of $T_c=26$~K superconductivity in
LaFeAsO$_{1-x}$F$_{x}$ has brought a new round of investigations
since iron is supposed to form magnetic order at low temperature.
Two classes of ferropnictide superconductors have drawn extensive
attention, i.e.\ doping of $Ln$FeAsO ($Ln =$~lanthanides) and
$A$Fe$_{2}$As$_{2}$ ($A =$~Sr, Ba, etc.) by holes or electrons.
The parent compounds of these two systems are suggested to have a
spin-density-wave (SDW) instability associated with the FeAs
layers in the temperature range between 130 and 200~K, that will
be gradually suppressed upon charge-carrier doping, and finally
the material becomes
superconducting.\cite{Kamihara08,Chen08a,Chen08b,Rotter08} By now
the relation of the antiferromagnetic spin-fluctuations to the
mechanism of superconductivity is not clear. Around \ts\ also a
structural transition is observed, where the symmetry changes from
tetragonal ($I4/mmm$) to orthorhombic ($Fmmm$).\cite{Kasinathan09}

Like in  BaFe$_2$As$_2$ and SrFe$_2$As$_2$, M\"o\ss{}bauer
spectroscopy and magnetic susceptibility studies of \efa\
revealed\cite{Raffius93,Shuai} that the magnetic transition due to
an antiferromagnetic (AFM) ordering of itinerant carriers in the
Fe sublattice that form a SDW; in the title compound the
temperature of the phase transition is approximately 190~K. The
peak in resistivity $\rho(T)$ supports this assignment; however,
the relation to the structural transition around the same
temperature is not solved by now.  Compared with other compounds
of the 122 family, the exceptional case in \efa\ is that in
coexistence to the SDW phase a second magnetic ordering takes
place in the localized Eu moments at $T_N=19$~K.\cite{Shuai}

While in BaFe$_2$As$_2$ Co- or Ni-substitution of Fe causes a suppression of the SDW and the appearance of a superconducting state,\cite{Sefat08,Li08} no superconductivity is found in the Eu analog EuFe$_{2-x}$Ni$_x$As$_2$;\cite{Ren08a} nevertheless, there seems to be some interaction between the Eu and Fe/Co/Ni subsystems, because the arrangement of the Eu moments turns to be ferromagnetic for $x>0.06$.\cite{Ren08a} Replacing Eu by K, for instance, supresses the SDW, and also substantially broadens the Eu order and shifts it down below 10~K; in spite of the short-range magnetic order, at $T_c=31$~K superconductivity is detected in Eu$_{0.5}$K$_{0.5}$Fe$_2$As$_2$.\cite{Jeevan08b,Gasparov09}

Applying pressure to \efa\ suppresses the SDW phase, too, but only some
onset of superconductivity was inferred at $T_c \approx 29.5$~K
where a drop in resistivity is detected above
2~GPa;\cite{Miclea08} the AFM ordering temperature of the
Eu$^{2+}$ moments is nearly unaffected by pressure. Very recently,
superconductivity was reported in \efa\ by partially substituting
As by isovalent P, inducing chemical pressure without
destroying the magnetic transition of the Eu$^{2+}$
moments.\cite{Ren08b} This agrees with predictions of magnetic
quantum criticality between an antiferromagnetic and paramagnetic
metal.\cite{Dai08} As a matter of fact, the ferromagnetic
interaction between the Eu$^{2+}$ moments (which is probably due
to RKKY interaction) is strengthened as a consequence of P doping
in a similar manner as observed in EuFe$_{2-x}$Ni$_x$As$_2$. Here
we investigate the influence of these different magnetic ordering
phenomena on the charge carrier dynamics in \efa\ in order to
elucidate the effect of spin fluctuations on the superconducting
ground state in ferropnictides, in general.

\section{Experimental Detail}
Single crystals of \efa\ were grown using FeAs as self-flux
dopants.\cite{Shuai} The platelets with a typical size of $4~{\rm
mm}\times 2~{\rm mm}\times 0.5~{\rm mm}$ have naturally flat
surface. The specimens from two different laboratories were
characterized by transport and susceptibility measurements; the
good agreement obtained in all results confirms the intrinsic
nature of the findings reported here. The temperature-dependent dc
resistivity was obtained by standard four-probe technique, using
silver paint as contacts. For the out-of-plane measurements, the
mechanic stability of the contacts during temperature sweeps and
inter-layer inclusions of presumably un-reacted precursor material cause
severe problems. Therefore particular care has to be taken to
obtain quantitatively reproducible results. \cite{remark1} The
magnetic susceptibility was measured down to 4~K by a Quantum
Design SQUID system at different magnetic fields up to 2 Tesla
applied parallel and perpendicular to the $ab$ plane.

The optical reflectivity was investigated by Fourier transform
spectroscopy with the electric field polarized in both the
$ab$-plane and the $c$-direction. The Bruker IFS 113v and IFS
66v/s interferometers utilized in our study cover the frequency
range from 30 to 12\,000~\cm. The experiments were performed at
various temperatures down to 10~K. As reference, gold was
evaporated onto the sample {\it in situ} and the measurements
repeated at each temperature. The $c$-axis infrared experiments
were performed using an infrared microscope Bruker Hyperion. For
the high-frequency extrapolation up to the ultraviolet, we used
room-temperature ellipsometric data (6000 - 30\,000~\cm) by a
Woollam variable angle spectroscopic ellipsometer. The
low-frequency extrapolation was done according to the dc
conductivity. The complex optical conductivity
$\hat{\sigma}(\omega)=\sigma_1(\omega)+{\rm i}\sigma_2(\omega)$
was calculated from the reflectivity spectra using Kramers-Kronig
analysis.\cite{DresselGruner02} It is worth to note, that we also
applied a simple Hagen-Rubens extrapolation for the low-frequency
reflectivity in $E\parallel ab$ polarization, and found that the
optical conductivity in the measured range does not depend on the
extrapolation. In the perpendicular direction, we extrapolated by
assuming a constant reflectivity.

In addition, we have
performed far-infrared reflection measurements in an optical
cryostat Oxford Spectromag 4000 where a magnetic field up to 5
Tesla can be applied perpendicular to the $ab$ plane. The
experiments were performed down to $T=2$~K with particular
emphasis on the magnetic ordering at $T_N=19$~K.

\section{Results}
\subsection{Electrical Resistivity}
\label{sec:resistivity}
\begin{figure}
    \centering
        \includegraphics[width=8cm]{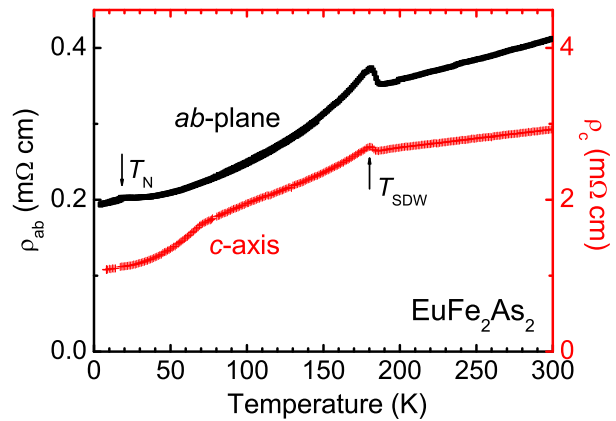}
    \caption{\label{fig:Fig1} (Color online) Temperature dependence of resistivity $\rho_{ab}$
    (black squares, left axis) in the $ab$-plane and $\rho_c$ (red crosses, right axis) in the perpendicular
    direction of single crystalline \efa. The spin-density-wave transition \ts\
    and the antiferromagnetic order of Eu at $T_N$ are indicated by arrows.}
    \end{figure}
\efa\ is a poor metal with a room temperature conductivity of only
$2.5\times 10^3~(\Omega{\rm cm})^{-1}$ which is more than two
order of magnitude lower compared to a normal metals, like iron or
gold. From the analysis of our optical data discussed below in
Sec.~\ref{sec:chargecarrierdensity}, we conclude that this is
mainly due to the reduced charge carrier
 density.
The electrical resistivity of several \efa\ crystals was measured
in the $ab$-plane and in $c$-direction. The results are plotted in
Fig.~\ref{fig:Fig1} as a function of temperature. When $T$ is
reduced below room temperature, the in-plane resistivity slowly
decreases as $\rho_{ab}\propto T$. This is similar to the behavior
reported for high-$T_c$ cuprates above $T^*$. A sharp upturn in
$\rho_{ab}$ is observed at $T_{\rm SDW}\approx 189$~K associated to
the SDW transition. With further cooling, the resistivity
decreases again, leading a peak in $\rho_{ab}$ around 189~K. The
rise in resistivity evidences the loss in density-of-states at the
Fermi energy upon entering the SDW phase. A similar peak is
observed in the perpendicular direction, $\rho_c(T)$, indicating
that the gap opens more or less isotropically over the entire
Fermi surface. However, the resistivity soon continues to drop
as the temperature is reduced further; this means that not
all  bands are affected by the SDW transition. Our conclusions are
in accord with previous
findings.\cite{Ren08a,Jeevan08b,Xiang08,Jeevan08a,Singh08,Yildirim09}

Compared to the linear resistivity behavior above \ts, that might
be influenced by fluctuations of the Fe spins, the slope of
$\rho(T)$ curve gets even steeper for $T<T_{\rm SDW}$, because the
scattering rate of carriers is reduced in the SDW state, where
some of the bands become (partially) gapped. Here the in-plane resistivity
follows a quadratic temperature dependence $\rho(T)\propto T^2$
all the way down to 30~K where it saturates at a constant value.
It seems unlikely that this behavior is due to electron-electron
scattering but might express scattering on magnetic excitations
where a $T^2$ behavior is predicted for a ferromagnet and $T^4$
for an antiferromagnet,\cite{Fournier93} with details strongly
depending on the dispersion relation.

At
$T_N=19$~K we observe a kink in $\rho_{ab}(T)$ because another
scattering channel freezes out due to the AFM ordering of the
Eu$^{2+}$ moments. Obviously, the energy scales of the SDW linked
to Fe and the AFM order in the Eu sublattice are different by an order of
magnitude, as indicated by the ordering temperature. The two phase
transitions also influence the electronic scattering very
differently: for $T<T_{\rm SDW}$ the scattering is strongly
reduced, while the change at $T_N$ is minor.

The resistivity measured along the perpendicular direction exceeds
$\rho_{ab}$ considerably: $\rho_c$(300~K)=3~${\rm m}\Omega{\rm
cm}$.\cite{remark1} Nevertheless, the temperature behavior
$\rho_c(T)$ closely resembles the in-plane properties; the
anisotropy ratio $\rho_{c}/\rho_{ab}\approx 8$ is almost
temperature independent. Hence, \efa\ exhibits a metallic behavior
of $\rho(T)$ in both directions. Although the 2D layered structure
is identified for the FeAs-based compounds as in cuprates, this
observation is distinct from high-$T_c$ cuprates which commonly
exhibit a different temperature dependences of the in-plane and
out-of-plane resistivity.\cite{Sadovskii} Our findings, of only
weak anisotropic in dc conductivity support the conclusion of
high-magnetic field measurements\cite{Altarawneh08,Yuan09}
indicating nearly isotropic superconductivity in
(Ba,K)Fe$_2$As$_2$.

The SDW of the Fe ions can clearly be seen in $\rho_c(T)$,
indicating its three dimensional nature. Similar observations have
been reported for the sister compound
BaFe$_2$As$_2$.\cite{XHChenBa} The AFM transition at $T_N=19$~K is
not seen in $\rho_c(T)$, in contrast to the in-plane resistivity.
This is somewhat surprising and infers that the magnetic order of
Eu$^{2+}$ ions is of short range and does not lead to a complete
three-dimensional order.

\subsection{Magnetic Susceptibility}
\begin{figure}
    \centering
        \includegraphics[width=7cm]{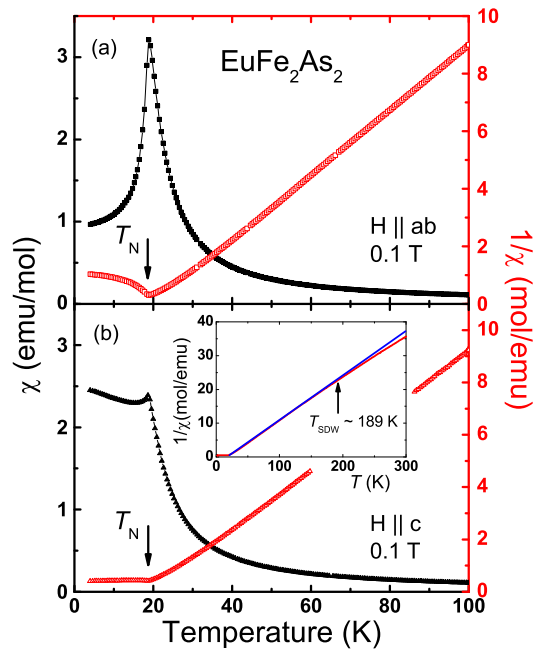}
    \caption{\label{fig:Fig2} (Color online) Magnetic susceptibility of of \efa\ for
    (a)~the $ab$ plane, and (b) the $c$-direction from $T=2$~K to 100~K (left axes).
    Corresponding to the right axes the inverse susceptibility  $1/\chi$ is plotted for both orientations.
    The inset shows the inverse susceptibility after subtracting the
    temperature independent terms for $H\parallel c$ (red-curve)
    in a wider temperature range ($2~{\rm K} < T < 300$~K);\cite{Shuai}
    the blue line indicates the Curie-Weiss fit up to 300~K.
    $1/\chi_{ab}$ (not shown) has the same trend as $\chi_c$ above 200~K.}
\end{figure}
Figure~\ref{fig:Fig2} exhibits the temperature dependence of the
magnetic susceptibility $\chi(T)$ of \efa\ measured for a magnetic field of
0.1 Tesla oriented both parallel and perpendicular to the $ab$
plane. Also shown is the inverse susceptibility $1/\chi_{ab}(T)$
and $1/\chi_c(T)$ which clearly proves that the paramagnetic
regime can be nicely described by the Curie-Weiss law. As
demonstrated in the inset of Fig.~\ref{fig:Fig2}, this behavior is
well observed over a very wide temperature range from 20~K to
200~K; here the magnetic response is isotropic. Slight deviations from $\chi(T)\propto 1/T$
are only observed for higher temperatures, i.e.\ above the SDW
transition which is identified as a tiny kink.

\begin{figure}
    \centering
        \includegraphics[width=6cm]{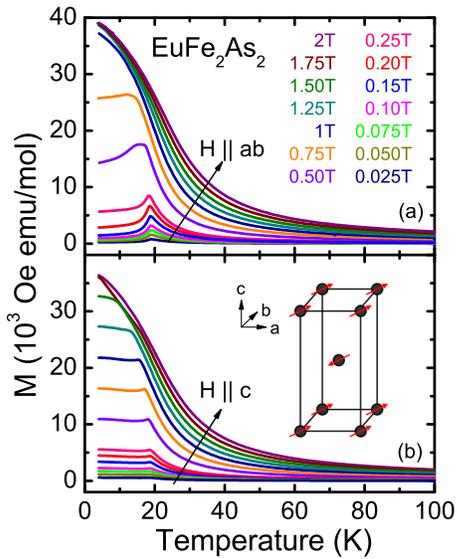}
    \caption{\label{fig:Fig3} (Color online) Temperature dependence of the magnetization of \efa\ under various fields up to 2~T; the development with magnetic field strength is indicated by the arrows. For both orientations, $M_{ab}$ and $M_c$ continuously  increase with rising magnetic field $H$,  as indicated by the black arrows. The inset shows a schematic diagram for the Eu magnetic structure. The long axis represents the $c$-axis; the red arrows indicate the spins of Eu which are not completely within the $ab$ plane.}
\end{figure}

At $T_N=19$~K, a distinct anomaly shows up in both orientations
that is ascribed to the ordering of the Eu$^{2+}$
moments.\cite{Shuai} It is concluded that the Eu$^{2+}$ spins
align ferromagnetically within the $ab$ planes, but
antiferromagnetically along the $c$-direction, as sketched in the
inset of Fig.\ref{fig:Fig3}. The upturn of $\chi_c(T)$ below
$T=18$~K indicates a metamagnetic transition of Eu in this
direction. The excellent quality of our samples are confirmed by
the good agreement of our results with the report by Jiang {\it et
al.}, \cite{Shuai} where also a more detailed description of
$\chi(T)$ can be found. It should be noted that although Fe and Eu
both carry magnetic moments, which is a unique behavior in
$A$Fe$_2$As$_2$ family, there is almost no coupling between both
subsystems; one reason is the large difference of energy scales
for local Eu ordering and itinerant Fe antiferromagnetism.
Nevertheless, the resistivity behavior and magnetoresistance
studied in Ref.~\onlinecite{Shuai} reflects that the charge
carriers are scattered by the Eu moments to some degree. In
addition, the absence of superconductivity in
EuFe$_{2-x}$Ni$_x$As$_2$ -- while it is present Ni doped
BaFe$_2$As$_2$ -- indicates that the magnetic state of Eu effects
superconductivity.\cite{Ren08a} In the same direction goes the
conclusion drawn from the Eu substitution by K, resulting in
$T_c=31$~K for
Eu$_{0.5}$K$_{0.5}$Fe$_2$As$_2$.\cite{Jeevan08b,Gasparov09}

Further investigations of the magnetization $M_{ab}(T)$ and
$M_c(T)$ under various magnetic fields are presented in
Fig.~\ref{fig:Fig3}. With increasing field up to 0.75~T, the AFM
transition in $ab$-plane gradually shifts to lower temperature.
For $\mu_0 H \geq$1~T, the metamagnetic transition is suppressed,
and $M_{ab}(T)$ tends to saturate below 10~K, indicating a full
ferromagnetic state of the Eu moments. In the case of $H\parallel
c$, a significantly higher external magnetic field ($\mu_0 H
>$1.5~T) is needed to fully suppress the AFM ordering of Eu
[Fig.~\ref{fig:Fig3}(b)]. This suggests that the Eu moments align
close to the $ab$-plane but still have a $c$-axis component [cf.\
the schematic diagram in Fig.~\ref{fig:Fig3}(b)].

\subsection{Optical Properties}
\begin{figure}
    \centering
   \includegraphics[width=7cm]{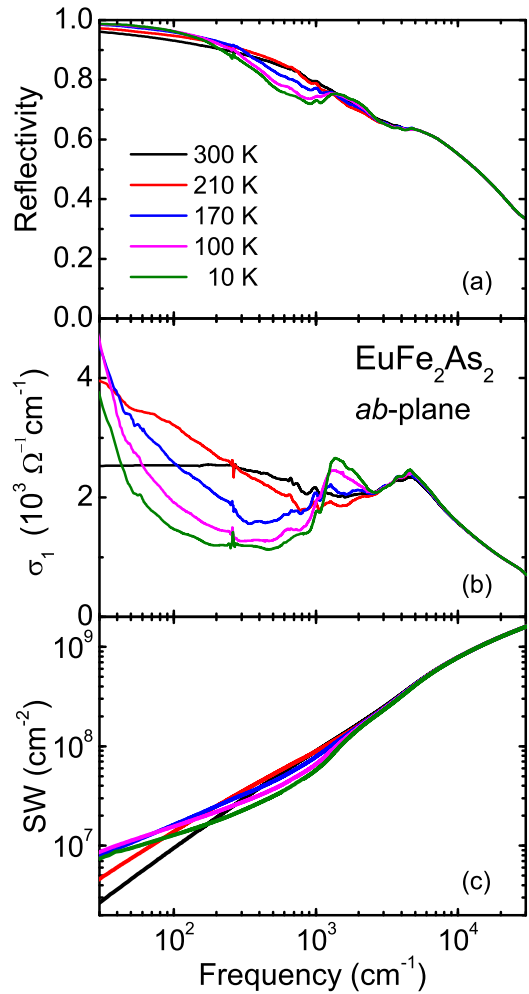}
    \caption{\label{fig:Fig4}(Color online) Optical properties of \efa\ in the $ab$-plane measured in a broad frequency range at different temperatures. (a)~The reflectivity $R(\omega)$ drops around 1250~\cm\ for $T=170$~K and lower since the SDW state is entered at $T_{\rm SDW}=189$~K. (b)~The corresponding optical conductivity indicates the development of the SDW gap around 1000~\cm; nevertheless there remains a strong background of excitations in the gap. (c)~The spectral weight calculated according to Eq.~(\protect\ref{eq:spectralweight}) is displayed in a double-logarithmic fashion.}
    \end{figure}

The upper panel of Fig.~\ref{fig:Fig4} shows the $ab$-plane
reflectivity $R(\omega)$ in the whole spectral range for some
selected temperatures as indicated. Above $T\approx 200$~K the
reflectance resembles a metal, although the plasma edge is not
clearly seen; a fact well-known from high-temperature and organic
superconductors.\cite{Basov05,Dressel04} This expresses the
poor metallic behavior, already seen in resistivity
(Sec.~\ref{sec:resistivity});
but also the overlap with interband transitions
leads to an overdamped plasma edge. At 1000~\cm\ the reflectance
is already as low as 80\%. Below the SDW transition, $T_{\rm
SDW}=189$~K, a drop in $R(\omega)$ is observed around 1250~\cm\
that is fully developed when $T$ reaches 10~K. The reflectivity is
strongly suppressed between 200 and 1000~\cm; but it rapidly
increases towards unity for $\omega\rightarrow 0$ because the
compound remains metallic for any temperature.

For all temperatures the in-plane optical conductivity spectra
$\sigma_1(\omega,T)$ of \efa\ show a broad peak in the
mid-infrared (5000~\cm) due to interband transitions
[Fig.~\ref{fig:Fig4}(b)].\cite{Singh08} This maximum gets slightly
stronger as the temperature is reduced, basically recovering the
spectral weight lost in the low-frequency region (see below). For
$T<T_{\rm SDW}$, the frequency region below 1000~\cm\ is depleted
and the spectral weight piles up around 1300~\cm. Although the
Drude contribution is appreciably reduced, it is always present:
indicating a partial gap opening of the Fermi surface. The
zero-frequency peak narrows upon cooling, because the decreasing
scattering rate of the charge carriers. This is in accord with the
observation from a falling $dc$ resistivity (Fig.~\ref{fig:Fig1}).

\begin{figure}
    \centering
   \includegraphics[width=7cm]{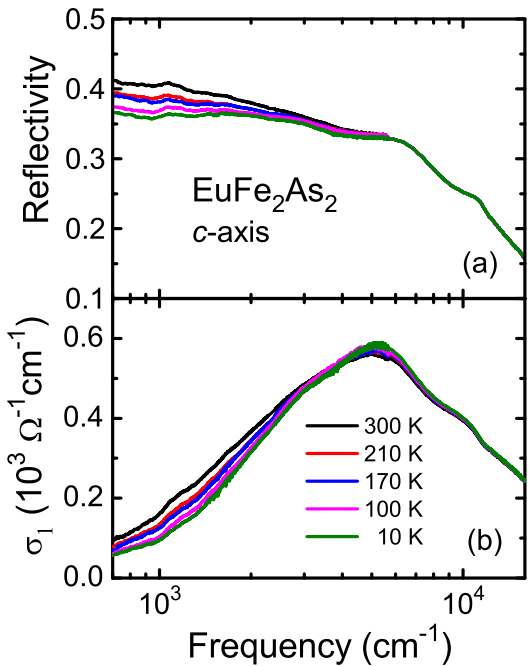}
    \caption{\label{fig:Fig5}(Color online) Optical properties of \efa\ measured at various temperatures
    with the electric field $E$ polarized in the $c$-direction.
    (a)~The mid-infrared reflectivity is much lower compared to the $ab$-plane reflectance and decreases even
    further when the temperature is reduced. (b)~The optical conductivity is dominated by the mid-infrared band around 5000~\cm\ ascribed to an interband transition.}
    \end{figure}

In the $c$-direction (Fig.~\ref{fig:Fig5}), the $R(\omega)$ shows
only 40\%\ reflectance in the mid-infrared and gradually falls at
high frequencies. On lowering the temperature, $R(T)$ decreases by
not more than 10\%. The optical conductivity exhibits a broad
maximum around 5000~\cm\ which only slightly varies with
temperature; it corresponds to the same interband transition
detected in the $ab$ plane. Although $\sigma_1(T)$ is reduced by a
factor of 2 in the range of 1000~\cm, when going from room
temperature to 10~K, the indications of a SDW gap are not as clear
as for the in-plane conductivity. Considerable changes are already
seen for $T>200$~K and the variation extends all the way up to
3000~\cm, i.e.\ well above the energy identified as the SDW gap by
the in-plane measurements. Thus we conclude that the formation of
the SDW state is not observed in the optical properties for
$E\parallel c$ in spite of the clear feature present in the dc
resistivity $\rho_c(T)$. In general, the substantial difference in
the optical properties of polarization $E\parallel c$ compared to
the in-plane results is in compliance with the dc anisotropy ratio
of 10, although we could not see indications of a Drude-like peak
in our limited frequency range.

\section{Analysis and Discussion}
\subsection{Charge Carrier Density}
\label{sec:chargecarrierdensity} As a first step, we  fit the
conductivity spectra of \efa\ by the Drude-Lorentz
model;\cite{DresselGruner02} this allows us to separate different
contributions. As an example, in Fig.~\ref{fig:Fig7}(a) the terms
are plotted for the low-temperature data
$\sigma_1(\omega,T=10~{\rm K})$. If the conductivity peak around
5000~\cm\ is modelled by a Lorentz term and ascribed to an
interband transition, the quasi-free-carrier parts remaining at
lower-frequencies yields a plasma frequency of approximately
14\,000~\cm\ at room-temperature. This value for \efa\ is in good
agreement with $\omega_p/2\pi c\approx 12\,000$~\cm\ obtained in
the sister compounds BaFe$_2$As$_2$ and SrFe$_2$As$_2$.\cite{Hu08}

Alternatively, we can calculate the spectral
weight\cite{DresselGruner02}
\begin{equation}
{\rm SW}(\omega_c) = 8\int_0^{\omega_c}\sigma_1(\omega)\,{\rm d}\omega = {\omega_p^2}=\frac{4\pi~ne^2}{m}
\label{eq:spectralweight}
\end{equation}
as a function of cut-off frequency $\omega_c$. As seen in
Fig.~\ref{fig:Fig4}(c), no obvious step or saturation evidences a
plasma frequency up to 30\,000~\cm; this is in accord with the
gradual decrease of the reflectivity $R(\omega)$ to higher
frequencies [Fig.~\ref{fig:Fig4}(a)]. Although the spectral weight
shifts to higher frequencies below the SDW transition, it is
basically recovered around 2500~\cm\, as can be seen from the
merger of the different curves in Fig.~\ref{fig:Fig4}(c). Hence,
the overall spectral weight remains conserved at any temperature.
The same behavior was observed in optical experiments of the Ba
and Sr analogues.\cite{Hu08,Pfuner08} To leave out the interband
transition at 5000~\cm\ and consider only the itinerant electrons,
we have chosen $\omega_c/2\pi c=2500$~\cm\ as suitable cut-off
frequency and then get 13\,800~\cm\ for the quasi-free-carrier
plasma frequency.

Using Eq.~(\ref{eq:spectralweight}) with $m=2m_0$ the
bandmass -- as obtained from our extended Drude analysis discussed below in Sec.~\ref{sec:extendedDrude} -- and $\omega_p/2\pi c \approx 13\,800$~\cm,
the carrier density is estimated to be
$n\approx 4.2\times 10^{21}~{\rm cm}^{-3}$,
significantly lower than typical for conventional metals.
If we convert $n$ to the number of carriers $N$ per unit cell
by using the low-temperature volume $V_{\rm cell}=0.3683$~nm$^3$ for $Z=4$
reported by M. Tegal {\it et al.},\cite{Tegal} we obtain $N
\approx 0.39$ electrons per formula unit.

\subsection{Spin-Density-Wave Gap}
\label{sec:spindensitywavegap}
\begin{figure}
    \centering
  \includegraphics[width=7.5cm]{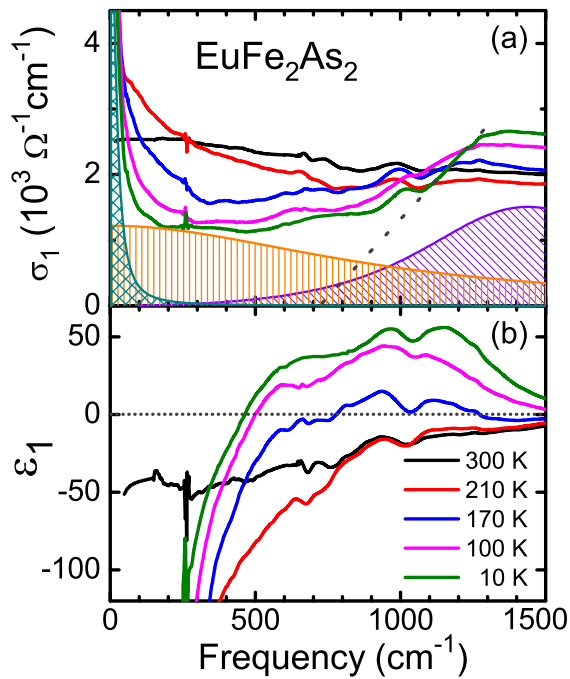}
    \caption{\label{fig:Fig7}(Color online) Low-frequency optical properties
    (up to 1500~\cm) of \efa\ at various temperatures for $E\parallel ab$.
    (a)~The optical conductivity can be decomposed in two Drude-like contribution
    (olive pattern and orange shaded), and an oscillator to mimic the SDW gap (violet shaded),
     as depicted for the 10~K data for instance. (b)~The frequency dependence of the dielectric
     constant exhibits a zero-crossing (${\rm d}\epsilon_1/{\rm d}\omega > 0$) as low as 500~\cm.
    }
    \end{figure}

Upon passing through the SDW transition, there is a strong
reduction of the optical conductivity below
1200~\cm\ as magnified in the linear-frequency presentation of
Fig.~\ref{fig:Fig7}(a). With decreasing temperature the peak above
the SDW gap grows and also the gap becomes slightly
larger; as expected from a mean-field transition. From a linear
extrapolation of the 10~K conductivity (dotted line), we can
estimate a gap value of $2\Delta_0\approx 750$~\cm, well above the
mean-field value of $2\Delta_0=3.53 T_{\rm SDW}\approx 470$~\cm. In
SrFe$_2$As$_2$ ($T_{\rm SDW} \approx 200$~K) Hu {\it et
al.}\cite{Hu08} identified two gaps with peaks at 500 and
1500~\cm; accordingly for BaFe$_2$As$_{2}$ ($T_{\rm SDW} = 138$~K)
they find the double peak features at lower energies (360~\cm\ and
890~\cm). The Eu compound investigated in the present study falls
right between with $T_{\rm SDW} = 189$~K, and the 1300~\cm\
maximum is likely to correspond to the reported high-frequency peak. However,
we cannot find clear indications for a low-energy gap. Around
600~\cm, a small step might be identified, in particular in the 100~K
spectrum, but since it smears out upon lowering the
temperature instead of getting more pronounced, we hesitate to
associate it with a second gap. Also the dielectric constant shown
in Fig.~\ref{fig:Fig7}(b) gives no evidence for a two-gap
structure. Interestingly, for BaFe$_2$As$_2$ Pfuner {\it et al.}
\cite{Pfuner08} report a pseudogap of 500~\cm, but they do not
observe changes at higher energies, in contrast to Hu {\it et
al.}\cite{Hu08,remark2} Multiple-gap features are not surprising
for a compound with many bands crossing the Fermi energy; similar
properties have been reported in MgB$_2$,\cite{Kuzmenko02} the model compound of a two-gap superconductor.

To estimate the spectral-weight shift upon opening of the SDW gap,
we identify an isobetic point $\omega_{i}\approx 1050$~\cm\ in the
optical spectra below which $\sigma_1(\omega)$ decreases as
$T<T_{\rm SDW}$ and above which the spectral weight piles up. As
can be seen in Figs.~\ref{fig:Fig4}(b) and \ref{fig:Fig7}(a), $\sigma_1(\omega_i)$ basically remains
unaltered with temperature. The relative shift in spectral weight
\begin{equation}
\Delta{\rm SW} =\frac{{\rm SW}(\omega_{i},T=300~{\rm K}) -{\rm
SW}(\omega_{i},T=10~{\rm K})}{{\rm SW}(\omega_{i},T=300~{\rm
K})}
\end{equation}
is approximately 35\%.

Supplementary information can be obtained from the dielectric
constant $\epsilon_1(\omega)= 1-4\pi\sigma_2(\omega)/\omega$ plotted in Fig.~\ref{fig:Fig7}(b). The zero-crossing of
$\epsilon_1(\omega)$ at $\omega_0/2\pi c=1270$, 1600 and 1780~\cm,
for $T=170$, 100 and 10~K, respectively, is an alternative method
to identify the SDW gap.

At all temperatures, there remains some zero-frequency
contribution that pulls the dielectric constant negative as
expected for a metal. For low temperatures, the zero-crossing with
positive slope (${\rm d}\epsilon_1/{\rm d}\omega > 0$) occurs
around $\tilde{\omega}_p/2\pi c
=\omega_{p,D}/\sqrt{\epsilon_1}\approx 460$~\cm, where
$\tilde{\omega}_p$ denotes the screened plasma frequency of the
zero-frequency contribution which shifts only little for $T\leq
100$~K. From Fig.~\ref{fig:Fig7}(b) we see that the dielectric
constant reaches $\epsilon_1\approx 55$ for $T=10$~K, which yields
$\omega_{p,D}/2\pi c\approx 3400$~\cm\ for the narrow Drude-like
contribution. This value agrees well with the one we get from the
Drude-Lorentz fit shown in Fig.~\ref{fig:Fig7}(a), where the
narrow Drude contribution has $\omega_{p,D}/2\pi c = 3150$~\cm\ at
$T=10$~K.

It is apparent from Fig.~\ref{fig:Fig7}(a), that there exists a
sizeable electronic background in the range between 200 and
700~\cm\ that stays at all temperatures. While at $T=300$~K this
contribution is not so clear, it starts to be well pronounced at
$T=210$~K and changes only little below 100~K. It should also be
noted that the background conductivity is present in all compounds
of this ferropnictide family; for instance, it is seen in the
superconducting Ba$_{0.55}$K$_{0.45}$Fe$_2$As$_2$, too, where the
magnetic order is suppressed by doping.\cite{Yang08} Hence we do
not related this term to the SDW transition. This electronic
background could be interpreted as a high-frequency tail of the
Drude-like contribution, appearing due to interactions of the
charge carriers with other excitations. Another possibility would
be to describe this feature by a broad Drude term with a width
around 1000~\cm\ [the orange shaded contribution in
Fig.~\ref{fig:Fig7}(a)]. Since the Fermi level is crossed by five
Fe $3d$ bands according to theoretical
studies,\cite{Jeevan08a,Xiang08} zero-frequency contributions
could stretch well above the far-infrared to account for
excitations in different Fe 3$d$ bands.

\subsection{Magnetic Field Dependence}
In order to obtain information on the magnetic scattering effects,
we have measured the in-plane optical reflectivity in the
frequency range $20- 700$~\cm\ when a magnetic fields was applied
up to 5~Tesla. In the far-infrared (up to 600~\cm) we
find an overall rise of the 10~K reflectivity  (not shown). For $\mu_0
H=1$~Tesla it is only 1\%, but increases to 5\%\ at 5~Tesla. The
reflectivity enhancement with magnetic field diminishes as the
temperature increases; above 30~K, no significant change is
observed. This infers that the scattering of the remaining
conduction electrons described by the broad Drude-term
in Fig.~\ref{fig:Fig7}(a)(orange shaded) is partially suppressed by the magnetic
field. This behavior does not change upon passing the magnetic order at $T_N=19$~K; which
implies that the ferrromagnetic order of the Eu moments taking
place above 1~Tesla has no appreciable influence on the optical
properties in this frequency range.

\subsection{Extended Drude Analysis}
\label{sec:extendedDrude}
\begin{figure}
\centering
\includegraphics[width=7cm]{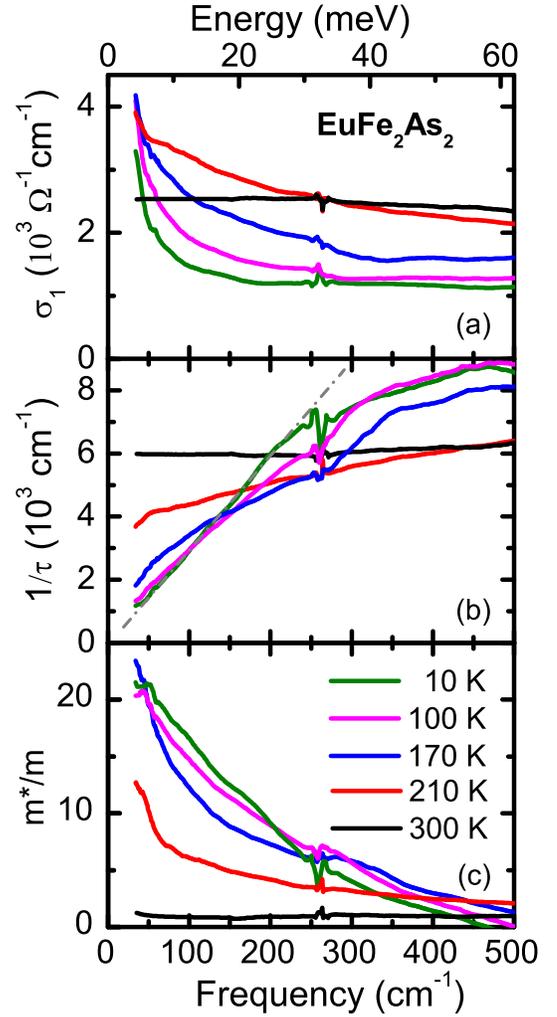}
    \caption{\label{fig:Fig8}(Color online) (a)~In-plane
    conductivity of \efa\ in the far-infrared spectral range.
    (b)~Frequency dependent scattering rate and (c)~frequency dependent
    effective mass as obtained from an extended Drude analysis of the optical spectra. The dash line in panel
    (b) indicates the linear behavior of $1/\tau(\omega)$.}
\end{figure}

A deeper insight into the low-energy excitations and the relevant
scattering mechanisms of \efa\ is obtained by performing an
extended Drude analysis of the conductivity spectra; here the
scattering rate $1/\tau(T,\omega)$ and effective mass
$m^*(T,\omega)$ are assumed to be frequency
dependent.\cite{DresselGruner02} This approach is commonly applied
to correlated electron systems like heavy fermions, organics and
high-temperature
superconductors:\cite{Dressel04,Dordevic06}
\begin{equation}
\hat{\sigma}(\omega)=\frac{\omega_p^2}{4\pi}
\frac{1}{\Gamma_1(\omega)-{\rm i}\omega(m^*(\omega)/m_{b})} \quad .
\label{eq:gen-drude2}
\end{equation}
Here $\Gamma_1(\omega)$ is the real part of a complex frequency
dependent scattering rate
$\hat{\Gamma}(\omega)=\Gamma_1(\omega)+{\rm i}\Gamma_2(\omega)$,
with the imaginary part related to the frequency dependent mass
$m^*/m_{b}=1-\Gamma_2(\omega)/\omega$ enhanced compared to the
bandmass $m_b$. From the complex conductivity we obtain
expressions for $\Gamma_1(\omega)$ and $m^*(\omega)$ in terms of
$\sigma_1(\omega)$ and $\sigma_2(\omega)$ as follows:
\begin{subequations}
\label{eq:extendedDrude}
\begin{eqnarray}
\Gamma_1(\omega)&=&\frac{\omega_p^2}{4\pi}
\frac{\sigma_1(\omega)}{|\hat{\sigma}(\omega)|^2}
\label{eq:gam-w}\\
\frac{m^*(\omega)}{m_{b}}&=&\frac{\omega_p^2}{4\pi}
\frac{\sigma_2(\omega)/\omega}{|\hat{\sigma}(\omega)|^2} \quad .
\label{eq:mstar-w}
\end{eqnarray}
\end{subequations}
The frequency dependent scattering rate $1/\tau(\omega) = 2\pi c
\Gamma_1(\omega)$ and mass $m^*(\omega)$ are plotted in
Figs.~\ref{fig:Fig8}(b) and (c) for various temperatures.

It is important to note that the absolute values of $1/\tau$ and
$m^*$ strongly depend on the chosen plasma frequency $\omega_p$ in
Eqs.~(\ref{eq:extendedDrude}); as mentioned above, this is not
that clear cut. Nevertheless, the frequency dependence is not
influenced by this renormalization and gives the insight into the
physics. From the room temperature spectra, using
$\omega_p=13\,800$~\cm, we can calculate an effective mass by
Eq.~(\ref{eq:spectralweight}) that  corresponds to the optical
bandmass $m_b\approx 2 m_0$. This is a reasonable value and
independent of frequency as expected for a Drude metal.
Accordingly, the scattering rate is constant at about
$1/\tau\approx 6000$~\cm. At $T=210$~K we already notice a gradual
increase of $m^*/m_b$ down to low energies
[Fig.~\ref{fig:Fig8}(c)]. The corresponding scattering rate
slightly increases with frequency due to the shaping up of the
narrow zero-frequency response, as plotted in Fig.~\ref{fig:Fig8}.

In the SDW state, some of the electronic bands are gapped,
implying that only a reduced number of carriers is available for
transport, as seen from the jump in $\rho(T)$ at \ts\ and the
opening of an optical gap below. Although the overall spectral
weight is eventually conserved, a considerable fraction  (as
discussed in Sec.IV-B) is shifted to energies above the gap.
Accordingly a reduced $\omega_p^*(T)$ should be used for the
generalized Drude analysis in Eq.~(\ref{eq:extendedDrude}) for
170, 100, and 10~K: $\omega_p^*/(2\pi c) \approx 12\,900$, 11\,900
and 11\,000~\cm, respectively. The enhancement of the effective
mass extends over a wide energy range and amounts up to
$m^*/m_b\approx 21$ for $T=10$~K as displayed in
Fig.~\ref{fig:Fig8}(c).\cite{remark3}

For $T<T_{\rm SDW}$, $1/\tau(\omega)$ is strongly suppressed at
low frequencies due to reduced phase space for scattering of
charge carriers upon opening of the SDW gap. This result agrees
with dc resistivity (cf.\ Sec.~\ref{sec:resistivity}) where we
concluded that the slope of $\rho(T)$ gets steeper due to the same
effect. We find a linear increase of $1/\tau(\omega)$ up to
approximately 200~\cm, as seen from Fig.~\ref{fig:Fig8}(b). This
evidences scattering on bosonic excitations, for instance
excitations of the spin density wave.\cite{Millis05} The slope
grows as the temperature decreases because the magnetic order is
completed further. The AFM ordering of the Eu$^{2+}$ ions at $T_N$
has little influence on the electronic scattering processes in
this energy range and seems to be confined to the in-plane dc
transport; which is not surprising since $\hbar\omega>k_B T_N$.
Below 500~\cm, the effective mass also starts to increase and
$m^*/m_b$ reaches more than 20 below $T_{\rm SDW}$, indicating
that the carriers are strongly interacting.

Here one should note that in the frequency range of our extended
Drude analysis, the conductivity is dominated by the response of
mobile carriers. In Fig.~\ref{fig:Fig8}(c) we can see that at
$m^*$ becomes negative at approximately 470~\cm. This is a clear
sign that the data above 470~\cm\ are dominated by
density-of-states effects, like energy gaps or interband
transitions; an interpretation in terms of scattering rate and
effective mass becomes meaningless.\cite{Millis05} Hence, to make
sure that the SDW gap does not mislead our analysis of the carrier
dynamics in terms of $1/\tau(\omega)$ and $m^*(\omega)$, we also
extracted all the high-frequency features, leaving only the
zero-frequency contributions for an extended Drude analysis. The
results obtained this way remain the same in the low-frequency
range as plotted in Fig.~\ref{fig:Fig8}; this strongly supports,
that our analysis procedure and interpretation is valid and
robust.

It is instructive to compare our findings with the spectra of
chromium, the canonical example of a three-dimensional SDW
system.\cite{Barker68,Basov02,Dordevic06} There also a gap opens
over some parts of the Fermi surface as a consequence of the SDW
ordering. Below \ts\ the Drude mode narrows and the low-energy
spectral weight is suppressed. From the extended Drude analysis a
suppression of the scattering rate in the SDW state was obtained
below 500~\cm\ and an overshoot for higher frequencies. Basov {\it
et al.} argue that the area above and below the SDW transition
should be conserved.\cite{Basov02}  Indeed, this is true if we
consider $1/\tau(\omega)$ in the whole frequency range. However,
the most important part of our extended Drude analysis is confined
to the narrow Drude-response (not analysed for Cr). Similar to the
case of Cr, this contribution becomes sharper for $T<T_{\rm SDW}$
due to the reduced phase space, described by a drop in
$1/\tau(\omega)$. The information is obtained from the particular
shape of the zero-frequency response and its temperature
dependence.

Yang {\it et al.} performed an extended Drude analysis of their
optical spectra on Ba$_{0.55}$K$_{0.45}$Fe$_2$As$_2$ after
subtracting the mid-infrared contribution described by a Lorentz
oscillator.\cite{Yang08} The low-temperature scattering rate
increases with frequency for $\omega/(2\pi c) < 400$~\cm, which
was interpreted as bosonic excitations -- most probable magnetic
fluctuations -- in support of our findings.

\subsection{Vibrational Features}
\begin{figure}
    \centering
        \includegraphics{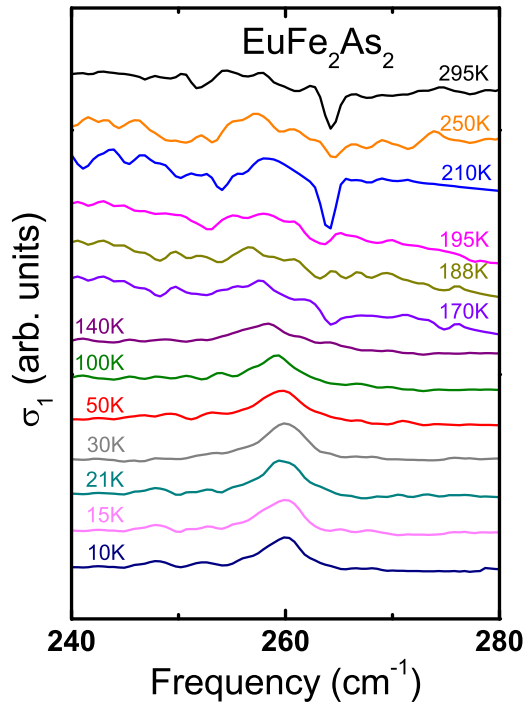}
    \caption{\label{fig:Fig6} (Color online) Temperature development of the conductivity in the region of the phonon mode of \efa. The different spectra are displaced for clarity reasons.}
\end{figure}
In our optical spectra displayed in Fig.~\ref{fig:Fig7}(a), a
phonon mode at 260~\cm\ is clearly observed for all the
temperatures. This E$_u$ mode involves the displacements of the Fe
and As ions and is of particular importance because it changes its
shape and position upon cooling. In Fig.~\ref{fig:Fig6} we magnify
this spectral region and plot the conductivity for various
temperatures. For $T\geq 170$~K, this phonon appears at 264.2~\cm\
as an anti-resonance. It can be followed down to $T=210$~K and
then becomes more and more obscured as $T$ decreases; below 170~K
the anti-resonance feature is basically absent. Instead a normal
phonon mode develops at lower frequency 257~\cm, which as a matter
of fact can be identified all the way up to 190~K. The mode
becomes stronger upon cooling down to $T=10$~K and hardens to
260~\cm.

\begin{table*}
\caption{\label{table1}Transport and optical parameters of \efa\
as obtained from our investigations on single crystals.
$\rho_{ab}$ and $\rho_c$ denote the room-temperature resistivity.
$\omega_p$ is the plasma frequency within the highly conducting
plane, from which the carrier density $n$ and the bandmass $m_b$
is obtained. From the room temperature scattering rate $1/\tau$,
the mean free path $\ell$ is evaluated, assuming a Fermi velocity
of $v_F=2\times 10^6$~m/s. $2\Delta$ is the value of the
zero-temperature SDW gap. \ts\ and $T_N$ refer to the ordering
temperatures of the SDW and AFM phase, respectively. At $T=10$~K
the effective mass related to the zero-frequency contribution is
enhanced by $m^*/m_b$.}
\begin{ruledtabular}
\begin{tabular}{cc ccccccccc}
$\rho_{ab}$ & $\rho_c$ & $\omega_p/(2\pi c)$ & $n$ & $m_b$ &
$1/\tau$ & $\ell$ & $2\Delta$
        & $T_{\rm SDW}$ & $T_N$  &$m^*$\\
(m$\Omega$\,cm) &(m$\Omega$\, cm) & (cm$^{-1}$) & (cm$^{-3}$) &$m_0$ &
(s$^{-1}$) & (nm) & (cm$^{-1})$
        & (K) & (K) & $m_b$ \\
\hline 0.41   & 3.0  & 13\,800 & $4.2\times 10^{21}$ & 2 & $1.8\times
10^{14}$ & 11 & 750 &189 & 19 & 21\\
\end{tabular}
\end{ruledtabular}
\end{table*}

The anti-resonance can be successfully fitted applying Fano's theory:\cite{FanoR, Menovsky} \begin{equation}
\sigma_1(\omega)\ = {\rm i}\sigma_0(q-{\rm i})^2({\rm i}+x)^{-1}
\label{eq:fano}
\end{equation}
where $\sigma_0$ is the background,
$x=(\omega^2-{\omega_T}^2)/\gamma\omega$ ($\gamma$ and $\omega_T$
are the linewidth and the resonant frequency, respectively) and
$q$ is the Fano parameter reflecting the degree of asymmetry of
the peak. The best description we get is $\omega_T=~263.4$~\cm\
and $q=-0.17$ for high temperatures. Such a low and negative value
of $q$ indicates the single level for vibration is overlapping
with the electronic background which is interacted.\cite{Menovsky}
For $T<170$~K, the fit yields $|q| \rightarrow \infty$ and the
Lorentz line shape is recovered, indicating a non-interacting case
appears upon the SDW gap opening. In conclusion, no new phonon
appears due to the structural phase transition, but the
significant modification of the vibrational feature around
260~\cm\ upon cooling is caused by the change of the electronic
background to which it is coupled. The coexistence of reminiscent
features of both limiting cases  -- the mode and the antimode --
in such a wide temperature range implies either strong fluctuation
effects, or inhomogeneities or phase separation in the vicinity of
the SDW transition.

The shape variation of this phonon mode by crossing the SDW
transition implies that the Fe electrons interacting with this
phonon are condensed to the gap feature. Our finding corresponds to
the conclusions from Raman scattering experiments on
CaFe$_2$As$_2$ and SrFe$_2$As$_2$,\cite{Choi08} which suggest the
variation in the phonon parameters is mainly caused by the change
of charge distribution within the FeAs plane and accordingly the
strength of the electron-phonon interaction. In the latest report
of first-principle calculations for CaFe$_2$As$_2$, it is also
suggested that modifying the chemistry of the Fe ion due to the
decrease of Fe-moment will decrease the Fe-As
interaction.\cite{Yildirim}

\section{Conclusions}
The magnetic, transport and optical properties of \efa\ single
crystals have been investigated parallel and perpendicular the
highly-conducting $ab$-plane. In Table~\ref{table1} we summarize
the parameters obtained. The anisotropy $\rho_c/\rho_{ab}\approx
8$ is basically temperature independent. From our optical data the
carrier density was estimated to $4.2\times 10^{21}~{\rm cm}^{-3}$
and the bandmass $m_b=2m_0$. The magnetic susceptibility is solely
determined by the localized magnetic moments of the Eu$^{2+}$ ions
with only little interaction to the Fe subsystem. The charge
carriers dynamics of \efa, on the other hand, is strongly affected
by the spin-density wave transition at $T_{\rm SDW}=189$~K when a
gap opens in the optical spectrum around 1000~\cm. A modification
of the Fe-As lattice vibration upon opening of the SDW gap
reflects an interaction with the electronic background that gets
much less pronounced in the SDW state. The remaining charge
carriers are strongly influenced by scattering at spin
fluctuations that can be modified by an external magnetic field.
The extended Drude analysis gives a linear dependence of the
scattering rate with frequencies at low temperatures. The
effective mass $m^*/m_b$ enhances by a factor of 21 at $T=10$~K.
Both parameters evidence the interaction of the low-energy charge
carriers with SDW excitations.

\begin{acknowledgments}
We thank J. Braun for ellipsometric measurements in the visible range.
We acknowledge help of and discussions with B. Gorshunov and V. I. Torgashev.
The work was partially supported by the Deutsche Forschungsgemeinschaft (DFG).
N.B. acknowledges support from the Alexander von Humboldt-Foundation.
N.D. is grateful for the support by the Magarete-von-Wrangell-Programm of
Baden-W\"urttemberg. The work at Zhejiang University was supported by NSF of China.
\end{acknowledgments}


\begin{thebibliography}{99}
\bibitem{remark1}From our measurements of a number of crystals from
different sources, the exact shape of $\rho_c(T)$ is sample dependent
at low temperature. For instance, the hump around 70~K shows up in different strength,
depending on the crystal, and might be related
to FeAs. We also found that samples like to cleave in flakes which
might infer a larger out-of-plane resisitivity than intrinsically
present.
\bibitem{remark2}This obvious discrepancy can only be explained by differences in sample quality. In our case of \efa\ the optical results on several crystals of different origin consistently show only one clear-cut gap around 1000~\cm.
\bibitem{remark3}Alternatively, we could consider only the zero-frequency
contribution, which has been modelled by the Drude term in
Fig.~\ref{fig:Fig7} and led to the zero-crossing of
$\epsilon_1(\omega)$ around 500~\cm. In that case the low plasma
frequency $\omega_{p,D}/2\pi c \approx 3400$~\cm\ yields a
zero-frequency effective mass of only $m_D^*/m = 4$. The latter
approach, however, seems to be inappropriate because it accounts
only for the excitations below 100~\cm\ but does not describe the
far-infrared behavior in its full extent.
\end{thebibliography}
\end{document}